\documentclass[usenatbib,useAMS,a4paper]{mn2e}

\usepackage{txfonts}
\usepackage{epsfig}
\usepackage{times}

\def\msun{{\rm M_{\odot}}}

\title [The abundance discrepancy problem]
{Can X-rays provide a solution to the abundance discrepancy problem in photoionised nebulae?}
\author[B. Ercolano]{B. Ercolano\\
Institute of Astronomy, Madingley Rd, Cambridge, CB3 0HA, UK}

\date{Submitted:}

\pagerange{\pageref{firstpage}--\pageref{lastpage}} \pubyear{}

\begin{document}

\def\lta{\mathrel{\spose{\lower 3pt\hbox{$\mathchar"218$}}
     \raise 2.0pt\hbox{$\mathchar"13C$}}}
\def\gta{\mathrel{\spose{\lower 3pt\hbox{$\mathchar"218$}}
     \raise 2.0pt\hbox{$\mathchar"13E$}}}
\def\Msun{{\rm M}_\odot}
\def\msun{{\rm M}_\odot}
\def\Rsun{{\rm R}_\odot}
\def\Lsun{{\rm L}_\odot}
\def\19{GRS~1915+105}
\label{firstpage}
\maketitle

\begin{abstract}
We re-examine the well-known discrepancy between ionic abundances
determined via the analysis of recombination lines (RLs) and
collisionally excited lines (CELs). We show that abundance variations
can be mimicked in a {\it chemically homogeneous} medium by the presence of
dense ($n_H \ga 10^4$ cm$^{-3}$) X-ray irradiated regions which present different ionisation and
temperature structures from those of the more diffuse medium they
are embedded in, which is predominantly ionised by extreme-ultraviolet radiation.
The presence of X-ray ionised   
dense clumps or filaments also naturally explains the lower
temperatures often measured from O~{\sc ii} recombination lines and
from the Balmer jump when compared to temperatures determined by
CELs. We discuss the implications for abundances                                            
determined via the analysis of CELs and RLs and provide a simple                      
analytical procedure to obtain upwards corrections for CEL-determined                 
abundance. 
While we show that the abundance discrepancy factor (ADF) and the
Balmer Jump temperature determined from observations of the Orion Nebula can
simultaneously be reproduced by this model (implying upward
corrections for CELs by a factor of 1.15), we find that the required X-ray
fluxes exceed the known Orion's stellar and diffuse X-ray budget, if we assume
that the clumps are located at the edge of the blister. We propose,
however, that spatially resolved observations may be used to
empirically test the model, and we outline how the framework developed
in this letter may be applied in the future to objects with better
constrained geometries (e.g. planetary nebulae). 


%
%
%
%
%
%
\end{abstract}


\section{Introduction}
The analysis of emission line spectra from photoionised gas
(e.g. in H~{\sc ii} region and planetary nebulae, PNe) often provides one of the
best method to determine elemental abundance in our own and other
galaxies. Accurate metallicities are fundamental to the
characterisation of many astrophysical system, and a number of empirical
techniques have been developed in order to extract this information from the
observation of nebular emission lines. Historically, collisionally
excited lines (CELs) provided the best options, thanks to their
brightness and their temperature and density diagnostic power. One of
the major drawbacks of CELs is their exponential dependence on
electron temperatures, which introduces uncertainties in the derived
abundances, particularly for high metallicity objects, where large
temperature gradients may occur (e.g. Stasi\'nska et al. 2005). 

Recombination lines (RLs), produced by electron cascades following recombination
of metal ions, are potentially more reliable abundance estimators,
since they have a much weaker (inverse) dependence on temperature and
are also only weakly dependent on density variations in the nebular
regime. However, due to their faintness (roughly $10^{-3}$ relative to
$H\beta$), they are only detectable in nearby objects and through very
deep observations. Nevertheless RLs from e.g. O~{\sc ii} and C~{\sc
  ii} are now routinely detected in many bright H~{\sc ii} regions in
the Galaxy and in the Magellanic clouds (e.g. Tsamis et al. 2003; Garcia-Rojas \& Esteban, 2007, and references therein)
, as well as in many PNe (Tsamis et al. 2004, 2008; Garcia-Rojas
et al. 2009; Wesson et al. 2005, 2008; Wang \& Liu, 2007; Krabbe \&
Copetti, 2006; Ercolano et al. 2003b, 2004). However, already over sixty
years ago, comparison of ionic abundances obtained by RLs and
CELs had revealed a systematic discrepancy such that RL-determined
abundances are always found to be higher than those determined from CELs
(Wyse 1942; Aller \& Menzel, 1945), these findings have been
confirmed for all 
observed nebulae. The  problem affects both H~{\sc ii} 
region abundances, with abundance discrepancy factors (ADF) of
order two (e.g. Garcia-Rojas et al. 2007, and references therein), and PNe,
where very large ADFs have been reported for some
extreme objects (e.g ADF $\sim$80 for Hf2-2, Liu et al. 2006).  This
is known as the {\it abundance discrepancy problem} and is probably
the major outstanding problem of nebular astrophysics (for further
discussion see Sections 5.11 and 9.5 of Osterbrock \& Ferland, 2006). 

The most important question is which set of abundances is more
reliable, a question that can only be answered when the cause of the
discrepancy is known. A number of solutions have been proposed, which include (i) temperature fluctuations, also known as the
$t^2$ paradigm (e.g Peimbert 1967), and (ii) hydrogen deficient inclusions
(e.g. Liu et al. 2000; Stasi\'nska et al. 2007). It is beyond the scope of this letter to
review the extensive research carried out by the supporters of (i)
or (ii), however it should be noted here that temperature
fluctuations in chemically homogeneous objects imply that CELs underestimate abundances and that RLs
provide the more accurate estimate. The presence of metal-rich
(hydrogen deficient) inclusions, implies instead that RLs are predominantly
emitted from these inclusions and therefore the abundance derived from
RLs is strongly affected by this gas-phase and it is not representative of
the bulk of the nebula. 

Both theories have limitations, and a large body of literature exist
in support and against one or the other. Problems with the $t^2$
paradigm include the fact that it cannot be used to explain ADFs
larger that a few, furthermore photoionisation models have so far
failed to reproduce the temperature fluctuations required
to explain even low values ADFs (e.g. Ercolano, Bastian \& Stasi\'nska,
2007). A problem with the metal-rich inclusions is in proving their
existence, a number of origins have been proposed for H~{\sc ii}
regions (e.g. Stasi\'nska et al. 2007) and PNe (e.g. Ercolano et al. 2004),
but no conclusive evidence has been brought forward.  

In this letter we investigate whether the solution to the abundance discrepancy
problem may lie in the existence of X-ray irradiated high density ($\ga
10^{4}$cm$^{-3}$) quasi-neutral clumps and filaments embedded in the
more tenuous EUV-ionised nebular gas. We show how this {\it chemically
homogeneous} two-phase medium may naturally account for typical H~{\sc
  ii} region ADFs, and discuss existing observational evidence in
support of our theory. We conclude however that the known X-ray
(diffuse and point) sources in H~{\sc ii} regions, fall short of the
requirements of our model. 

In Section~2 we derive a relation of the ADFs to the relative
electron temperature and ionisation structure of the two phases and
their relative ionised masses. In Section~3 we discuss the
photoionisation process in a quasi-neutral gas irradiated by X-rays and
present results of photoionisation modelling of plane parallel slabs
using typical conditions, as inferred from available
observations. In Section~4 we further discuss our results and their
implications for the determination of gas abundances from nebular
emission lines in the Orion Nebula. Our conclusions are summarised in Section~5.

\section{Abundance discrepancy factor in a two-phase medium}

We consider the case of a two phase medium comprising high density
($\ga$10$^4$~cm$^{-3}$) clumps ionised by X-ray
radiation embedded in a lower
density component (e.g. $<$1000 cm$^{-3}$), predominantly ionised by
the extreme-ultraviolet (EUV) radiation field from young OB stars,
clusters or complexes. In the next section we will briefly review
photoionisation processes in quasi-neutral X-ray irradiated gas, but for
the sake of the current discussion it is sufficient to define the ion
fraction of ion $A^i$ of species $A$ as $x(A^i)$, such that
the total number of 
$A^i$ ions in a region with neutral hydrogen number density $n_H$ and volume
$V$ is given by N$(A^i) = x(A^i)\cdot(A/H)\cdot n_H \cdot V$, where
$A/H$ is the total abundance of element $A$ by number relative to
hydrogen. We further define $j_{RL}$ as the temperature and density dependent
emissivity of a given emission line produced by
recombination of $A^i + e^- \rightarrow A^{i-1} + h\nu$.   

The observed $RL/H\beta$ flux ratio ($\Re_{RL}$) emerging from such a region is given by 

\begin{equation}
\Re_{RL} = 
\frac{A}{H}\frac{(j_{RL}^X\cdot x(A^{i})^X\cdot n_H^X \cdot n_e^X \cdot V^X)
  +(j_{RL}^E\cdot x(A^{i})^E \cdot n_H^E\cdot n_e^E \cdot V^E)}
{(j_{H\beta}^X\cdot EM^X)
  +(j_{H\beta}^E\cdot EM^E) }
\label{e:big}
\end{equation}

\noindent where the $E$ and $X$ superscripts denote
physical quantities for the EUV and X-ray ionised regions (regions E and X),
respectively, and we have defined the ionised hydrogen emission measure as

\begin{equation}
EM = x(H^{+})\cdot n_H \cdot n_e \cdot V
\end{equation}

\noindent We note that, since we are considering a chemically
homogeneous medium, $A/H$ is the same in both regions. 

We define the constant $\varepsilon$ which represents the ratio of the ionised
hydrogen emission measures in the two regions: 

\begin{equation}
EM^E = \varepsilon \cdot EM^X
\label{e:a}
\end{equation}

RL emissivities carry an approximately inverse dependence on electron temperature and
are also weak functions of electron density, therefore we define $\iota$
and $\iota'$ such that, 

\begin{equation}
\frac{j_{RL}^X}{j_{RL}^E} \simeq \frac{j_{H\beta}^X}{j_{H\beta}^E} \simeq \iota 
\label{e:b}
\end{equation}

\noindent Substituting Equations~\ref{e:a} to \ref{e:b} into
Equation~\ref{e:big} and rearranging we obtain

\begin{equation}
\left(\frac{A}{H}\frac{x(A^i)}{x(H^+)}\right)_{RL}^{obs} = \frac{\Re_{RL}}{j} =
\frac{A}{H}\left[\frac{x(A^{i})^X}{x(H^{+})^X}\cdot\frac{\iota}{\iota+\varepsilon} +
\frac{x(A^{i})^E}{x(H^{+})^E}\cdot\frac{\varepsilon}{\iota+\varepsilon}\right]
\label{e:d}
\end{equation}

\noindent where we have defined $j = \frac{j_{RL}^E}{j_{H\beta}^E}
\simeq \frac{j_{RL}^X}{j_{H\beta}^X}$.

$\frac{\Re_{RL}}{j}$ is what an observer would use to determine  $\left(\frac{A}{H}\frac{x(A^i)}{x(H^+)}\right)$ from RLs, and therefore we have
called it $\left(\frac{A}{H}\frac{x(A^i)}{x(H^+)}\right)_{RL}^{obs}$. 

An observer measures $ADF(A^i)$ by comparing
$\left(\frac{A}{H}\frac{x(A^i)}{x(H^+)}\right)_{RL}^{obs}$ to
$\left(\frac{A}{H}\frac{x(A^i)}{x(H^+)}\right)_{CEL}^{obs}$, 
where the CEL analysis is based on the flux ratio of CEL lines
(e.g. $[OIII]\lambda 5007, 4959~{\AA}$) to $H\beta$ at a given T$_e$. As shown in the
next two sections, temperatures in a
quasi-neutral X-ray heated region (X region) are lower than those in the ionised EUV
heated region (E region) and 
therefore we can assume that the CEL flux is dominated by region $E$
(due to the exponential dependence on electron temperature of the CEL
emissivities). However 
$H\beta$ is likely to be produced in both regions, in proportions
dictated by the parameters $\iota$ and $\varepsilon$, defined in
Equation~\ref{e:a}. Namely, 

\begin{eqnarray}
\nonumber I(H\beta)^{tot} &= (j_{H\beta}^X\cdot EM^X)
  +(j_{H\beta}^E\cdot EM^E)\\
                &= j_{H\beta}^E\cdot EM^E\cdot \left(\frac{\iota+\varepsilon}{\varepsilon}\right).
\end{eqnarray}

\noindent It follows that

\begin{equation}
\left(\frac{A}{H}\frac{x(A^i)}{x(H^+)}\right)_{CEL}^{obs} = \frac{A}{H}\frac{x(A^{i})^E}{x(H^{+})^E}\cdot\frac{\varepsilon}{\iota+\varepsilon}
\label{e:cel}
\end{equation}

\noindent and 
\begin{eqnarray}
ADF(A^i) &=
\nonumber \left(\frac{A}{H}\frac{x(A^i)}{x(H^+)}\right)_{RL}^{obs}/\left(\frac{A}{H}\frac{x(A^i)}{x(H^+)}\right)_{CEL}^{obs}\\       
         &= W \cdot \frac{\iota}{\varepsilon} + 1
\label{e:final}
\end{eqnarray}

\noindent where, for conciseness, we have defined the parameter 

\begin{equation}
W = \frac{x(A^{i})^X}{x(H^{+})^X} / \frac{x(A^{i})^E}{x(H^{+})^E}.
\end{equation}

The upward correction for CEL-determined $A^i$ abundances, $\chi(A^i)$, can be
 obtained from equation~\ref{e:cel}: 
\begin{equation}
\chi(A^i) = \frac{\iota+\varepsilon}{\varepsilon}.
\label{e:corrcel}
\end{equation}

\section{Ionisation structure of dense gas irradiated by a tenuous X-ray field}

\begin{table*}
\label{t:t1}
\begin{center}
\begin{tabular}{lccccccccc}
\hline
\hline 
Model & log($F_X$)         &  log($n_H$)  & $<T_e(H^{+})>$ & $<x(H^{+})>$ &  $<x(O^{2+})>$ &$\iota^a$   &  $ \varepsilon^{b}$ &   $\chi (CEL)^{c}$ & T(BJ)$^{d}$ \\
  & $[$phot/s/cm$^{-2}]$  &  $[$cm$^{-3}]$ & $[$10$^3$K$]$ &  $[$10$^{-3}]$&  $[$10$^{-3}]$ &                 &            &                 & $[$10$^3$K$]$ \\ 
\hline                                                                                                                                          
F6N4  &      6          &     4           &  1.2          &  0.71       &  0.90      &          6.91          &  38.4     &   1.18   &     5.1      \\
F6N5  &      6          &     5           &  1.3          &  0.22       &  0.22      &          6.27          &  27.5     &   1.23   &     4.8      \\
F6N6  &      6          &     6           &  1.2          &  0.05       &  0.06      &          7.24          &  26.5     &   1.27   &     4.2      \\
F7N4  &      7          &     4           &  1.6          &  2.5        &   4.1      &          5.27          &  37.9     &   1.14   &     6.1      \\
F7N5  &      7          &     5           &  1.5          &  0.83       &  0.97      &          5.62          &  28.8     &   1.19   &     5.4      \\
F7N6  &      7          &     6           &  1.4          &  0.23       &  0.26      &          6.02          &  29.8     &   1.20   &     5.1      \\
F8N4  &      8          &     4           &  2.3          &  9.2        &  18.2      &          3.66          &  31.8     &   1.12   &     6.9      \\
F8N5  &      8          &     5           &  2.0          &  3.3        &  4.6       &          4.21          &  25.7     &   1.16   &     6.2      \\
F8N6  &      8          &     6           &  1.8          &  0.86       &  1.2       &          4.68          &  28.6     &   1.16   &     6.0      \\
F9N4  &      9          &     4           &  4.5          &  36         &  57        &          1.87          &  13.0     &   1.14   &     7.7      \\
F9N5  &      9          &     5           &  3.2          &  12         &  20        &          2.63          &  19.2     &   1.14   &     7.2      \\
F9N6  &      9          &     6           &  2.8          &  4.1        &  6.2       &          3.01          &  20.0     &   1.15   &     6.9      \\
\hline                                                                                          
\hline
\end{tabular}
\end{center}

\caption[]{Model parameters and results for X-ray irradiated
  quasi-neutral slabs. $^a$ Estimated for Orion assuming $\frac{x(O^{2+})^E}{x(H^+)^E} = 0.57$ (Balwin et al. 2001), $T_e^X = <T_e(H^{+})>$ and 
  $T_e^E = 8430~$K (Garcia-Rojas et al. 2007). The emissivities of the O{\sc ii}~V1 multiplet are from Storey (1994) and Smits
  (1991). $^{b}$Inferred from Equation~\ref{e:final} for the Orion
  ADF$\simeq$1.4 (Garcia-Rojas et al. 2007). $^{c}$Inferred from
  Equation~11. $^{d}$Inferred from Equation~\ref{e:bj}; the observed T$_e$(BJ) in Orion is 7900$\pm$600~K. }
\end{table*}     
            
Multiwavelength studies of star-forming regions like Orion and 30~Dor have
revealed a highly fragmented structure where warm molecular gas
and ionised gas are intermixed (Poglitsch et al. 1995). 
 The high density clumps and filaments are opaque to EUV photons, but
they are maintained at a low level of ionisation by X-rays, which can
penetrate much larger neutral
hydrogen columns (e.g. $10^{22}$cm$^{-2}$ at $\sim$1~keV). The
ionisation structure of these regions is very different from that of a
typical H~{\sc ii} gas. X-rays can photoionise the inner shells of
atoms and ions, from which Auger multi-electron ejection may
follow. This process couples non adjacent stages of
ionisation. Furthermore high energy radiation may also eject suprathermal electrons which then produce secondary ionisations (e.g. Xu
\& McCray, 1991). 


We have used the photoionisation code {\sc mocassin} (Ercolano et al. 2003, 2005,
2008) to calculate the ionisation and temperature structure of dense
slabs irradiated by a thermal X-ray spectrum. Table~1 lists the model
parameters for the illustrative cases presented here. 
The slabs are irradiated using a synthetic spectrum obtained for an
X-ray temperature of log($T_X$) = 6.8~K using the
procedure detailed in Ercolano et al. (2008). 
We use ``solar'' abundances from Grevesse \& Sauval (1998) with the
exception of C,N and O abundances which are taken from Allende-Prieto
et al. (2002), Holweger (2001) and Allende-Prieto, Lambert \& Asplund
(2001), respectively. 
The density of clumps and filaments in real nebulae is likely to vary,
however their density must be 
large enough to become opaque to EUV radiation. Poglitsch et al. (1995)
suggest densities as high as $10^6$~cm$^{-3}$, in agreement with observations of
molecular gas in ionised regions by other authors who inferred densities ranging
from $10^4$cm$^{-3}$ to $10^6$cm$^{-3}$. Therefore here we experiment with
slabs of densities in this range, adjusting the length of the slabs
such that they are truncated at a column of $10^{21}$cm$^{-2}$. 

For the range of densities and X-ray fluxes covered by our
models, we find that very low electron temperatures can be
produced in the slabs while a low level of ionisation is still
maintained. Table~1 lists the values of
$<T_e(H^+)>$, $<x(H^{+})>$ and $<x(O^{2+})>$, defined as

\begin{equation}
<T_e(A^i)> = \frac{\int N_{e} x(A^i) T_{e} \, dV}{\int N_{e} N(A^i) \,
dV}
\end{equation}
\begin{equation}
<x(A^{i})> =  \frac{\int N_e\,x(A^i) \, dV}{\sum_i\int N_e N(A^i) \, dV}
\end{equation}

\noindent The X-ray irradiated slabs show very different
characteristics from those of EUV irradiated H~{\sc ii} gas, which has generally much higher
electron temperatures (roughly 7000 to 12000~K) and $<x(O^{2+})>$ to
$<x(H^+)>$ ratios always less than unity (Stasi\'nska, 1982). 

However before proceeding to our discussion of the implication of our
models for the ADF in Orion, it is worth mentioning a few caveats. 
(i) The results presented here are for low-ionisation/neutral gas and
do not include a chemical network for reactions typical of X-ray
dominated regions (XDR; e.g. see Meijerink \& Spaans, 2005). As a
consequence heating and cooling channels relevant to these regions
have also been ignored. However our models are restricted to low
column densities ($10^{21}cm^{-2}$) where molecular hydrogen is unlikely to be present.
(ii) Due to the large neutral hydrogen and helium abundances, charge
exchange plays an important role in the ionisation balance of heavy
elements. The accuracy of the ionisation structure predicted by our models
finally depends on the uncertainties in the charge exchange rates
adopted (Ferland et al. 1997).
(iii) The dense clumps are also bathed in a far-ultraviolet radiation (FUV)
field. While this will not affect the ionisation structure of the gas,
it may affect its temperature, principally via photoelectric emission
from the surface of grains. (iv) Finally, grain depletion may also be
more important in dense clumps and in the photon dominated
region, modifying ADFs for carbon and oxygen. In these cases there may
be variations within an H~{\sc ii} region depending on how much of
each phase the line of sight passes through. The blister geometry of
Orion may be a good example. These processes are not included in the
current models.  

While, in view of these limitations, we caution that the temperatures reported
here are not to be considered accurate, we stress however that the
'real' temperatures will still be factors of several below the
EUV-heated gas temperatures, and therefore the general conclusions of
our work remain unaffected. Nonetheless this underlines the need for
further modeling which takes into account of XDR physics and FUV irradiation. 

\section{Application to the Orion Nebula}
Abundance discrepancy factors (ADFs) have been measured in a number of
H~{\sc ii} regions and cluster around the value of two
(e.g. Garcia-Rojas \& Esteban, 2007 and references therein
). Due to its proximity, Orion is one of the best studied
regions and has a reported mean ADF value of 1.4 (e.g. Garcia-Rojas et
al. 2007).  Mesa-Delgado et al. (2008) presented deep long slit spectra at
several positions in the Orion Nebula. They found that $ADF(O^{2+})$
was roughly constant along the slit positions, except at the location
of Herbig-Haro objects where the largest increases were
recorded. In the framework of the clumpy X-ray + diffuse EUV (X-E) model, 
the enhanced $ADF(O^{2+})$ at these locations is due to X-rays from
the Herbig-Haro shock front impinging on cold neutral material (Pravdo
et al. 2001). Also in agreement with this model, enhanced values of $ADF(O^{2+})$ were also
reported by these authors in the region closer to $\theta^1~Ori~C$
(which emits $\sim$60\% of the entire X-ray flux in the Trapezium). 

The physical conditions that can create a {\it mean} ADF value of 1.4
in Orion are now explored in the context of the X-E model.  
We use $\iota = \frac{j_{O^{2+}V1}^X}{j_{O^{2+}V1}^E}$, $T_e^X =
<T_e(O^{2+})>$ and $W =
\frac{<x(O^{2+})>^X}{<x(H^+)>^X}/\frac{x(O^{2+})^E}{x(H^+)^E}$  and assume
values for the EUV ionised region as determined by
observations (e.g. Garcia-Rojas et al. 2007) and detailed
photoionisation modeling (Baldwin et al. 1991) which give
$\frac{x(O^{2+})^E}{x(H^+)^E} = 0.57$ and   
  $T_e^E = 8340\pm130~$K. The emissivities of the O{\sc ii}~V1 multiplet are
calculated using the data of Storey (1994) and Smits (1991). We use
the above to determine $\iota$ and $W$ from our models
(see Table~1) and using these in Equation~\ref{e:final} we find the
values of $\varepsilon$, the ratio of ionised hydrogen emission
measures in regions E and X, needed to produce an ADF of 1.4. From the
values of $\varepsilon$ and $\iota$ thus calculated, 
$\chi(O^{2+})$, the upward correction for the CEL-determined $O^{2+}$
abundance, can also be easily determined (Equation~11) and is
listed in Table~1. The $O^{2+}$
abundance in the E-region, in this case, can be obtained
from the CEL value after an upward correction by factors of
1.12 to 1.27. 

The values of $\varepsilon$ shown in Table~1, are encouraging as they show
that the ionised emission measure required for the $X$ region is always
significantly smaller than that in the $E$ region. An observational
constraint on $\varepsilon$ can be teased out from a 
comparison of temperature measurements from the Balmer Jump ($T_e(BJ)$) and from
collisionally excited lines ($T_e(CEL)$). Balmer Jump (BJ) measurements will contain
contributions from both the X and the E regions, while CELs are
expected to be dominated by the E region. Starting from this
assumption one can derive the following relation:

\begin{equation}
T_e(BJ) = \left(\frac{\varepsilon + \iota^{5/3}}{\varepsilon +
  \iota}\right)^{-3/2} \cdot T_e(CEL)
\label{e:bj}
\end{equation}
\noindent where we have used the fact that the BJ to H11 line
ratio is proportional to temperature to the power of -2.3
(e.g. Liu et al. 2001). 

From the equation above and the values of $\varepsilon$ and $\iota$
inferred from our models and Orion-type E-region conditions and
ADF values, we have calculated the predicted $T_e(BJ)$ (last column of
Table~1), which we can compare with the observed value of 7900$\pm$600
K (Garcia-Rojas et al. 2007). The comparison shows that models with an
X-ray flux of $10^9$phot/sec provide the best match, which implies an
upward correction to the CEL abundance of 1.14.



\subsection{Origins of the X-ray flux}
H~{\sc ii} regions present a number of sources of X-ray radiation. 
In the Orion Nebula we can simplistically list interacting stellar winds
as a source of the diffuse component (e.g. Guedel et al. 2008) and
young stars emitting X-rays as point sources (e.g. Feigelson et al.
2005). The observed diffuse X-ray source in the Orion is quite small
(L$_X$~=~$5.5\times10^{31}$erg/s; Guedel et al. 2008) and cannot produce
fluxes in the range discussed in Section~3. However the stellar wind
itself carries a much larger mechanical luminosity. The wind from
$\theta^1$Ori~C alone has L$_{\rm mech}$~=~7$\times$10$^{35}$erg/s. If
all of its energy were transformed into 300eV photons upon impact with the
clumps, this would correspond to a photon flux of 1.3$\times$10$^9$ phot/s for
clumps located at the wall of the blister (roughly 0.1 pc distance
from $\theta^1$Ori~C, Baldwin et al. 1991; Schiffer \& Mathis, 1974),
matching the requirements of the X-E model. Considerations of radiative timescales, however, make the above 
rather unlikely. Indeed the low density stellar wind will form a bow
shock around the clump, but, due to the extremely low number
density of the wind the radiative timescales will be very long and
only a very small fraction of the impacting mechanical flux will be
radiated in the X-ray at the shock-front. If the clumps are embedded in the H~{\sc ii} region gas the wind would have to first interact with the ionised material, further complicating matters.

The ONC stars radiate approximately 5.7$\times$10$^{33}$erg/sec in the X-ray 
(Feigelson et al. 2005), which corresponds to a photon luminosity of
roughly 3.5$\times$10$^{42}$phot/s (assuming 1~keV photons given that
the stellar spectra are much harder than shock-produced emission). A
photon flux of $\sim$10$^9$phot/s, which gives the best fit to Orion's ADF
and BJ temperature (see Table~1), would  require the clumps to be
placed at a distance of $\sim$1.7$\times$10$^{16}$cm from the source,
which is about 15 times smaller than the estimated separation of the
blister's wall from its ionising source (Baldwin et al. 1991; Schiffer
\& Mathis, 1974). At the edge of the blister the photon flux would be
only $\sim$3$\times$10$^6$phot/s; as shown in
Table~1, X-ray fluxes of order 10$^6$phot/s produce a very cold
X-region and as a consequence underestimates the BJ temperatures. 

\section{Conclusions}

In this letter we have explored the possibility that X-ray ionised
dense clumps embedded in lower density EUV-ionised gas may explain the
discrepancy between ionic abundance determinations based on the
analysis of RLs and CELs in H~{\sc ii} regions. While the model can,
in principle, simultaneously reproduce the observed ADF and BJ
temperature in the Orion Nebula, the X-ray flux requirements do not
match the observed X-ray emission. 

Independent from our current findings in the Orion Nebula, future
observational studies may be used to empirically probe the X-E model
by searching for a correlation between local ADF variations with
(measured or expected) local X-ray fields. If notwithstanding our
pessimistic analysis above, this model were found to hold, the main
implications for emission line abundance determinations would be that
both CEL- and RL-determined abundances would require adjustments. CEL
corrections are readily available using the simple method derived in this work. 
Furthermore, unlike the t$^2$ paradigm, the X-E model predicts that
the error on the abundances derived from CELs is always smaller and
easier to correct for than that on RL-determined abundances. This
stems from the fact that RLs include contributions from both the X and
E regions which have very different temperatures and ionisation
structures. The CELs on the other hands are mainly produced in the E
region and error on those is only introduced by the fact that the
H$\beta$ line which is used in the abundance analysis contains
contributions from the two regions. However this is easy to correct
for by following the procedure outlined in this paper. 

In this letter we have mainly concentrated on the H~{\sc ii} regions, however, larger ADFs have been recorded in some PNe 
(e.g. Wesson et al. 2003, 2008; Ercolano et al. 2004; Liu et al. 2006). In recent years 
the large difference in the magnitude of the ADFs for PNe and H~{\sc
ii} regions has led several authors to conclude that different effects
  may be at play in these systems (e.g. Garcia-Rojas et
  al. 2007). It is beyond the scope of this paper to discuss the
  abundance discrepancy problem in PNe in detail, however we speculate
  that also in these system cold, quasi-neutral X-ray irradiated dense
  gas may the culprit for the high ADFs reported.  
X-ray emission has
  indeed been observed in a number of PNe (e.g. Chu et al. 1997;
  Guerrero et al, 2005; Montez \& Kastner, 2009), and it is well known
  that many PNe contain a dense equatorial disk, optically thick to EUV radiation
  (e.g. Lester \& Dinerstein, 1984). High density clumps have also been directly imaged
  in many PNe, like the Helix Nebula (O'Dell et al. 2004) and are likely to
  be present in many other objects. 
Furthermore, Tsamis et al. (2004) found that for a given PN the CEL
C/O and N/O ratios  were very similar to the RL C/O and N/O ratios,
respectively. 
This lends further support to chemically homogeneous models like the
X-E model, where CELS 
and RLs come from regions with the same heavy element abundances. 
PNe with their more easily constrained geometry, may indeed, be a better
laboratory to test the feasibility of this scenario.

\section{Acknowledgments}
We warmly thank (alphabetically) M. Barlow, N. Bastian, C. Clarke,
J. Drake, A. Glassgold, M. Guedel, W. Henney, D. Hollenbach,
J. Raymond, G. Stasi\'nska, P. Storey and R. Wesson for their critical
assessment of our work.

\end{document}